\documentstyle[12pt,fleqn,epsf]{article}
\def\ie{\hbox{\it i.e. }}        
        
\def\etal{\hbox{\it et al.}}

\def\abs#1{\left| #1\right|}

\def\ztautau{\ \hspace*{-1.4mm} \raisebox{-1.2mm}{$_{_Z}$} _{\tau\tau}\ }
\def\nuww{\ \hspace*{-1mm} _{\nu} \raisebox{.7mm}{$_{_{WW}}$} \ }
\def\nuws{\ \hspace*{-1mm}  _\nu \raisebox{.6mm}{$_{_W}$} _{\sigma}\ }
\def\nusw{\ \hspace*{-1mm}  _{\nu\sigma} \raisebox{.6mm}{$_{_W}$} \ }
\def\tfz{\ \hspace*{-1mm} _{\tau\Phi} \raisebox{.7mm}{$_{_Z}$} \ }
\def\tzf{\ \hspace*{-1mm} _{\tau} \raisebox{.7mm}{$_{_Z}$} _\Phi \ }
\def\wnunu{\ \hspace*{-1mm}  \raisebox{-.3mm}{$_{_W}$} _{\nu\nu} \ }

\def\beq{\begin{equation}}
\def\eeq{\end{equation}}
\def\bea{\begin{eqnarray}}
\def\eea{\end{eqnarray}}

\def\np#1#2#3{    {\it Nucl. Phys. }{\bf #1} (19#2) #3}
\def\pl#1#2#3{    {\it Phys. Lett. }{\bf #1} (19#2) #3}
\def\pr#1#2#3{    {\it Phys. Rev. }{\bf #1} (19#2) #3}

\def\prl#1#2#3{    {\it Phys. Rev. Lett. }{\bf #1} (19#2) #3}

\def\zp#1#2#3{    {\it Zeit. f\"ur Physik }{\bf #1} (19#2) #3}

\def\mxfigura#1#2#3#4{
  \begin{figure}[hbtp]
    \begin{center}
      \epsfxsize=#1
      \leavevmode
      \epsffile{#2}
     \end{center}
    \caption{#3}
    \label{#4}
  \end{figure} }
\def\prepnuma{FTUV/94-43}
\def\prepnumb{IFIC/94-38}
\def\titol{The tau weak-magnetic dipole moment}
\def\autora { J. Bernab{\'e}u, G.A. Gonz{\'a}lez-Sprinberg, M.Tung and J.
Vidal}
\def\adressaa{Departament de F{\'{\i }}sica Te{\`o}rica,
Universitat de Val{\`e}ncia, \\ and IFIC, Centre Mixt Univ. Valencia-CSIC\\
E-46100 Burjassot (Val\`encia), Spain\\August 1994}
\def\resum{
We calculate  the prediction for the anomalous weak-magnetic form factor
of the tau lepton at $q^2=M_Z^2$  within the Standard Model.
 With all particles on-shell, this is a electroweak gauge invariant
 quantity. Its value is
 $a_\tau^w (M_Z^2)= - \;(2.10 + 0.61\, i) \times 10^{-6}$.
We
show that the transverse and normal components of
the single-tau polarization of tau pairs
produced at $e^+e^-$ unpolarized collisions are
sensitive to the real and absorptive parts of the anomalous weak-magnetic
dipole moment of the tau. The sensitivity one can
achieve at LEP in the measurement of this dipole moment is discussed.}

\def\today{\ifcase\month\or
 January\or February\or March\or April\or May\or June\or July
 \or August\or September\or October\or November\or December\fi
 \space\number\year}
\def\firstpage{\begin{titlepage}
\baselineskip 0.50cm \null
\vspace*{-1cm}
 \hfill \prepnuma\\
\null \hfill \prepnumb\\
\baselineskip 0.75cm
\vskip 4.cm
\begin{center}
{\Large \bf \titol}
\vskip 2.cm
\baselineskip 0.6cm
{\bf \autora}
\vskip 1.cm
{\it \adressaa}
\vskip 2.cm
{\sc abstract}
\end{center}
\baselineskip 0.60cm
\begin{quotation} \resum
 \end{quotation}

\end{titlepage}\baselineskip 0.75cm}

\begin{document}
\firstpage
\setcounter{page}{2}

\def\tendeix#1#2#3{
      \stackrel{#2\rightarrow #3}{\longrightarrow} #1}
%-------------------------------------------------------
%
%--------------------BODY----------------
%
%--------------------------------------------------------
\section{Introduction}
The anomalous magnetic dipole moments of the electron and  muon provide very
precise tests of quantum electrodynamics. The Standard Model
predictions can also be confronted with these properties. In
particular, $\tau$-physics still offers an open window to surprises.

In this paper we calculate, within the Standard Model, the one loop
anomalous weak-magnetic
moment (AWMM) of the $\tau$ lepton at the energy scale of the $Z$,
and find a way to measure it.
We show that for $e^+\, e^- \longrightarrow
\tau^+ \tau^-$ unpolarized scattering at the $Z$-peak, the
transverse (within the collision
plane) and normal (to the collision plane)
 single $\tau$ polarizations are very sensitive to the
real and imaginary parts of the
anomalous weak-magnetic ($a_{\tau}^w(M_Z^2)$) dipole form factor,
respectively. We  discuss
our results  as a background in order to separate this signal in the search
for new physics.

Polarization measurements are accessible for the $\tau$ by means
of the energy and angular distribution of its decay products.
The angular distribution of the
$\tau$-polarization, measured at LEP \cite{lep},
 contains separate information \cite{jadach,alemany,pover}
on both the average
polarization, measuring parity violation in the $Z-\tau^+-\tau^-$ vertex,
and the
$Z$-polarization, measuring parity violation in the $Z-e^+-e^-$
vertex.
The information on   the $Z-e^+-e^-$ vertex is also
available from the cross section asymmetry for longitudinally
polarized beams \cite{sld}, as measured by SLD.
This
allows a test of neutral current universality and, within the
Standard Model,
a precise determination of $\sin^2\theta_w$. No other component
of the
single $\tau$-polarization is allowed in the Standard Model
for unpolarized beams, in the limit of zero-mass fermions.

For the $e^+e^- \longrightarrow \tau^+\tau^-$ process at the
$Z$-peak,
the spin density matrix of the produced $\tau$ pairs
has single $\tau$-polarization terms that translate
into the energy and angular distribution of the decay products.
One can  try to
 isolate a weak-magnetic
dipole moment
term, looking for observables sensitive to this property in
the spin density
matrix of the  $\tau$ pairs. This kind of studies was done
 by the Heidelberg group \cite{nachtmann,bernreuther,overmann}
 in order to isolate the real and absorptive parts of a weak-electric
 dipole moment from spin correlations in
 $e^+e^- \longrightarrow \tau^+\tau^-$ decay products.
The $\tau$ weak-magnetic moment was also
investigated in \cite{stiegler} by looking to
forward and backward transverse asymmetries.

In Ref.\cite{nos} it has been demonstrated
 that the single transverse and normal $\tau$
polarization terms contain separate information about the (real part of the)
weak-magnetic and weak-electric dipole moments, respectively.

In this paper, with the help of the
angular distribution of the hadronic decay products acting as analyzers
  of the
spin components \cite{tsai,wag,rouge,martin},
we  explicitly construct asymmetries that can be used as
observables in the measurement of the
anomalous weak-magnetic form factor at
$q^2=M_Z^2$.
In particular, we show how  the transverse and normal components
 of the  single
$\tau$-polarization  provide
information about  the real and imaginary parts of the
weak-magnetic  form factor
 $a_{\tau}^w(M_Z^2)$.

In section 2 we  calculate the Standard Model prediction for the one
loop correction  to the $\tau$
anomalous weak-magnetic form factor at $q^2=M_Z^2$.
This magnitude measures the weak-magnetic coupling
$Z - \tau - \tau$. We  show that its value is essentially determined
 by the ratio of the square of the fermion mass
to the weak-boson mass. Furthermore, it is governed by both kinds of
diagrams
present and not present in the photon vertex case.
A very tiny dependence on the Higgs mass
is obtained.

In section 3 we show that $a_{\tau}^w(M_Z^2)$ can be measured by  using
the transverse and normal polarization of single $\tau$'s.
We extend our previous study on the dipole moments of Ref.\cite{nos},
so as to include the absorptive part of AWMM.  We find that the normal
polarization of single $\tau$'s is
sensitive to the absorptive part of AWMM.
 Finally, we discuss the sensitivity that can be achieved at LEP in the
measurement of this form factor. The appendix contains some of the
definitions and formulas we have used in the computation.
\section{Anomalous weak-magnetic moment}

In this section we calculate
 the Standard Model prediction
 for the anomalous
weak-magnetic form factor ( $ a^w_\tau(M_Z^2) $ )
 of the $\tau$ lepton at $q^2=M_Z^2$.
  This is an order $\alpha$
electroweak radiative
correction  to the weak magnetic moment.
The matrix element of the vector neutral current coupled
to the $Z$ is written, using Lorentz covariance, in the form
\beq
\bar{u}(p_-) \,V^{\mu}(p_-,p_+)\, v(p_+)
= \bar{u}(p_-) \left[ \frac{v(q^2) \gamma^\mu}{2 s_w c_w}+i
\frac{a^w_\tau(q^2)}{2 m_\tau}
\sigma^{\mu\eta} q_\eta\right] v(p_+)
\label{mu}\eeq
where $q=p_-+p_+$,  $e$ is the proton charge and
$s_w$, $c_w$
are  the weak mixing  angle sine and cosine,
respectively. The first term $v(q^2)$ is the Dirac
 vertex (or charge radius) form factor
 and it is present at tree level with a value
 $v(q^2)=\frac{1}{2}-2\, s_w^2$, whereas the second form factor
  only appears
due to quantum corrections.
Only the on-shell vertex with $q^2=M_Z^2$ is entitled to be
electroweak gauge invariant in the Standard Model.
$a^w_\tau$ can have contributions from both new
physics or electroweak radiative corrections  to the Standard Model.
We  calculate the leading Standard Model contribution to $a^w_\tau$.
One-loop contributions are formally of order
$\alpha$, but the magnitude of each diagram is in fact also
governed by the weak-boson
 or Higgs mass-factors like
 $\frac{m_\tau^2}{M_{Z}^2}$ or $\frac{m_\tau^2}{M_{\Phi}^2}$.
\mxfigura{13cm}{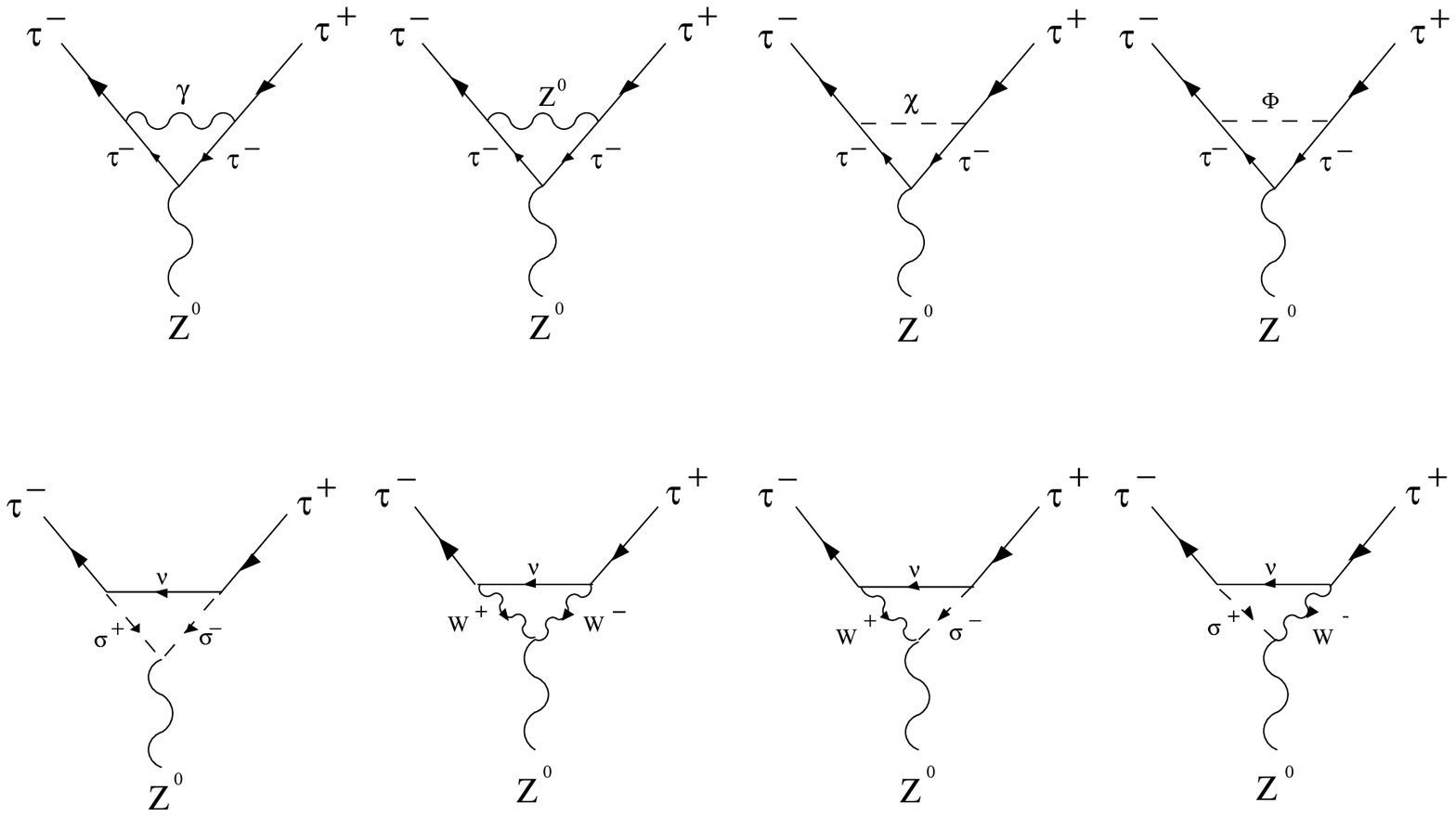}{Contributing Feynman
diagrams to $a^w_\tau$
in the t'Hooft-Feynman gauge  that are also present
for the  anomalous magnetic
moment (photon vertex) in the electroweak theory.}{Figure 1}
We compute the AWMM
in the t'Hooft-Feynman gauge, where no ambiguities in the
finite parts are present \cite{fuj}.
In principle, there are 14  diagrams to compute,  6 of which
 are not present in the photon vertex case. The  eight diagrams
that have a photon analogue
are shown in figure 1, and the  new ones are shown in figure 2.
\mxfigura{12cm}{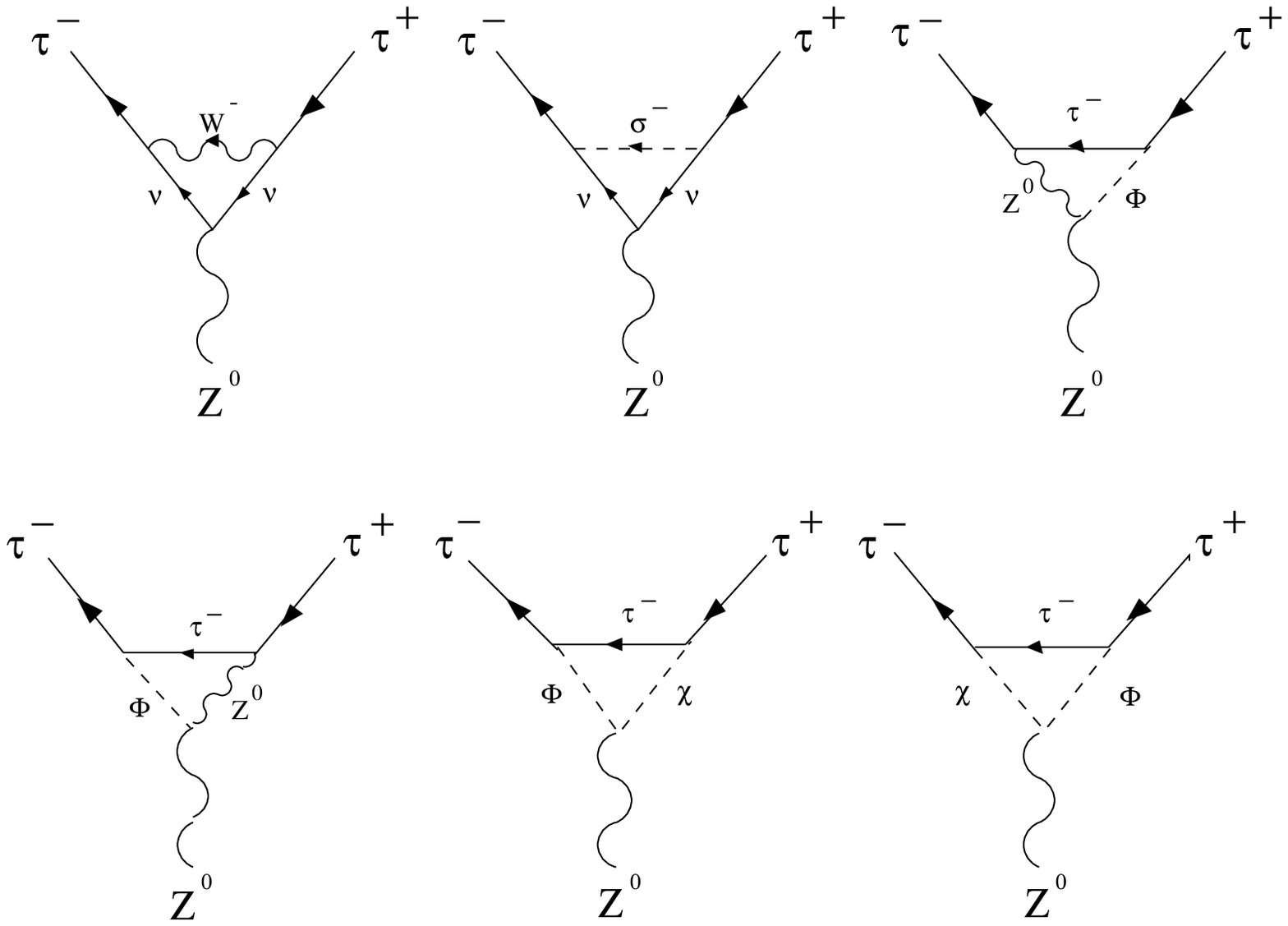}{Contributing Feynman
diagrams to $a^w_\tau$
in the t'Hooft-Feynman gauge  that are not present
in the  anomalous magnetic
moment case.}{Figure 2}
All  contributions are written as:
\beq
a_{ABC}=\frac{\alpha}{4  \pi} \;\;\frac{m_\tau^2}{M_Z^2}\;\,\sum_{ij}
c_{ij} {I_{ij}}^{ABC}
\label{a}\eeq
where $A$, $B$ and $C$ are the particles circulating in the loop, counting
clockwise in the diagrams from the particle between the two fermion lines,
$c_{ij}$ are
coefficients, and ${I_{ij}}^{ABC} \equiv
I_{ij}(m_\tau^2,q^2,m_\tau^2,m_A^2,m_B^2,m_C^2)$
 are scalar, vector or tensor functions defined
in  Eqs.(\ref{point}) and (\ref{3point}) of the Appendix.

When computing the diagrams we  only select the
tensor structure related
to the AWMM
(which is finite and needs no renormalization),
and we  also verify the vector current conservation
 as a check of our expressions
 (there is no induced $(p_-+p\prime)^\mu$ term in Eq.(\ref{mu})).
The external lines are on the mass shell, \ie $p_-^2=m_\tau^2\,$,
$p_+^2=m_\tau^2$ and $(p_-+p_+)^2=M_Z^2$ respectively.
We denote by $\sigma^\pm$ the charged  non-physical Higgs
and by $\chi$ and $\Phi$ the neutral non-physical and physical ones.
Some of the diagrams with  the
propagation of Higgs or would-be Goldstone bosons particles
are suppressed by extra ($\frac{m_\tau^2}{M_{Z,\Phi}^2}$) terms
in such a way that the $a_{\chi\tau\tau}, a_{\Phi\tau\tau},
a_{\sigma\nu\nu}$ and $a_{\sigma\sigma\nu}$ contributions to
$a^w_\tau$ are negligible.
 Diagrams in
which the  Higgs and the neutral would-be Goldstone boson particles
 couple to the $Z$ only contribute to the axial form factor
 and not to the magnetic moment
($a_{\tau\Phi\chi}=a_{\tau\chi\Phi}=0$).
In the following we give the different leading
contributions, where the notation
is self-explanatory:
\bea
\hspace*{-0.6cm}a_{\gamma\tau\tau}=\frac{\alpha}{4  \pi}
\frac{m_\tau^2}{M_Z^2} \;\frac{4 v M_Z^2}
{ s_w c_w} \;[ I_{10}
+I_{22}-I_{21}]^{\gamma\tau\tau }
\nonumber\eea
\bea
\hspace*{-0.6cm}a\ztautau =
\frac{\alpha}{4  \pi}
\frac{m_\tau^2}{M_Z^2} \;\frac{v M_Z^2}
{ s_w^3 c_w^3}\;
 [-4 a^2 I_{00}+(v^2+11 a^2) I_{10}
+(v^2+3 a^2) (I_{22}-I_{21})]^{\ztautau }
\nonumber\eea
\bea
\hspace*{-0.6cm}a\nuww =\frac{\alpha}{4  \pi}
\frac{m_\tau^2}{M_Z^2}\; \frac{ M_W^2}
{ s^3_w c_w} \;[ I_{10}+2 I_{21}-2 I_{22}]^{\nuww }
\nonumber\eea
\bea
\hspace*{-0.6cm}a\nuws =-
\frac{\alpha}{4  \pi} \frac{m_\tau^2}{M_Z^2} \;\frac{ M_Z^2}
{2  s_w c_w}  \;{I_{10}}^{\nuws }
\nonumber\eea
\bea
\hspace*{-0.6cm}a\nusw =-
\frac{\alpha}{4  \pi} \;\frac{m_\tau^2}{M_Z^2} \;\frac{ M_Z^2}
{2  s_w c_w}  \;{I_{10}}^{\nusw }
\nonumber\eea
\bea
\hspace*{-0.6cm}a\tfz =
-\frac{\alpha}{4  \pi} \;\frac{m_\tau^2}{M_Z^2} \;\frac{v M_Z^2}
{2  s_w^3 c_w^3}  \;{I_{10}}^{\tfz }
\nonumber\eea
\bea
\hspace*{-0.6cm}a\tzf =
-\frac{\alpha}{4  \pi} \;\frac{m_\tau^2}{M_Z^2} \;\frac{v M_Z^2}
{2  s_w^3 c_w^3}  \;{I_{10}}^{\tzf }
\nonumber\eea
\bea
\hspace*{-0.6cm}a\wnunu =\frac{\alpha}{4  \pi}
\frac{m_\tau^2}{M_Z^2} \;\frac{ M_W^2}
{ s^3_w c^3_w} \;[ I_{00}-3 I_{10}
-I_{22}+I_{21}]^{\wnunu }
\label{lead}\eea
where $v=-\frac{1}{2}+2 s^2_w$ and $a=-\frac{1}{2}$ are the vector and
axial vector $Z-\tau^+-\tau^-$ couplings.
The order of magnitude of each diagram
is given basically by the weak boson mass scale:
 all  are of the same order given by
$\frac{\alpha}{4  \pi} \frac{m_\tau^2}{M_Z^2}$.

The ${I_{ij}}^{ABC}$ functions were analytically computed
in terms of  dilogarithm functions, and checked with a numerical
integration for  the $m_\tau \rightarrow 0$ limit.
 Some details are given in the Appendix. We then obtain that the
numerical contribution of each diagram is:
\bea
a_{\gamma\tau\tau}=-\frac{\alpha}{4  \pi} \frac{m_\tau^2}{M_Z^2}
\;(1.32 - 0.52\,  i) \simeq (3.12 - 1.23\, i) \times 10^{-7}
\nonumber\eea
\bea
a\ztautau =\frac{\alpha}{4  \pi} \frac{m_\tau^2}{M_Z^2}
\;(0.17 + 0.08\, i) \simeq (3.92 + 1.88\, i) \times 10^{-8}
\nonumber\eea
\bea
a\nuww =\frac{\alpha}{4  \pi} \frac{m_\tau^2}{M_Z^2} \;(- 7.07 )
\simeq -1.68 \times 10^{-6}
\nonumber\eea
\bea
a\nuws = \frac{\alpha}{4  \pi} \frac{m_\tau^2}{M_Z^2} \;0.45
\simeq  1.06 \times 10^{-7}
\nonumber\eea
\bea
a\nusw = \frac{\alpha}{4  \pi} \frac{m_\tau^2}{M_Z^2} \;0.45
\simeq  1.06 \times 10^{-7}
\nonumber\eea
\bea
a\tfz = a\tzf =-\frac{\alpha}{4  \pi}
\frac{m_\tau^2}{M_Z^2} \;(0.07\;;\;0.03\;;\;0.02)
\simeq - (0.15\;;\;0.07\;;\;0.04) \times 10^{-7}
\nonumber\label{num}\eea
\beq
 a\wnunu =\frac{\alpha}{4  \pi} \frac{m_\tau^2}{M_Z^2}
\;(- 4.11 - 2.12\, i)
\simeq -\; (0.974 + 0.502\, i) \times 10^{-6}
\eeq
where the  values between parenthesis for $a\tfz = a\tzf$
correspond to $\frac{M_\Phi}{M_Z}=1,2,3$ respectively.

Finally, the value of the computed  AWMM is
\beq
a_\tau^w (M_Z^2)= - \;(2.10 + 0.61\, i) \times 10^{-6}
\label{anom}\eeq
The Higgs mass
only modifies the real part of this result less than a 1\%,
 from
the value $-\,2.12 \times 10^{-6}$ to $-\,2.10 \times 10^{-6}$ for
$1 < \frac{ M_\Phi}{M_Z} < 3$, and in Eq.(\ref{anom})  we have
chosen $M_\Phi=2 M_Z$.

To our knowledge,
no other calculation of this value has been reported up to now. In
Ref.\cite{queijeiro} a formal presentation of Feynman amplitudes
 in the unitary gauge is done, but
no explicit computation is made. There it is argued that
 the procedure would be legitimate if a rigorous computation
in a renormalizable
gauge, as we have done, is made.
We observe that the form factor has an absorptive part of
the same order as the real one.
We should point out that, contrary to
 the well
known electroweak anomalous magnetic moment,
this absorptive part is due to
the fact that we compute on the $Z$ mass
shell $q^2=M_Z^2$, not $q^2=0$.
In fact, one expects a non-vanishing imaginary part coming from unitarity.
We have also shown  that the main contribution
comes from the diagrams with $W$-exchange.
 Finally we
 would like to stress that this on shell result is gauge invariant,
 as this form factor
 is a linearly independent Lorentz matrix part of the contribution to
the physical $Z \longrightarrow \tau^+\tau^-$ decay.

\section{Observables related to the AWMM}

In this section we show that  the $a^w_\tau$  form factor
can be measured by
 observables related to the single $\tau$ polarization.
Keeping only up to linear terms in the spin and in the weak
dipole moments and
neglecting terms proportional to the electron mass,
the tree level  $e^+\, e^- \longrightarrow
\tau^+ \tau^-$ cross section at the $Z$-peak can be written as:
\begin{equation}
\frac{d \sigma}{d  \Omega_{\tau^-}}=
\frac{d \sigma^{0}}{d  \Omega_{\tau^-}}
+\frac{d \sigma^{S}}
{d  \Omega_{\tau^-}}
\label{cross1}
\end{equation}
where
the first term
collects the  spin independent terms, whereas the second one
takes into account the linear terms in  the spin:
\bea
\frac{d \sigma^{S}}{d \Omega_{\tau^-}}&=&
\frac{\alpha^2\beta }{ 128 s_w^3 c_w^3}\frac{1}{\Gamma_Z^2}
\left[\  (s_-+s_+)_x X_++  (s_-+s_+)_y Y_++
  (s_-+s_+)_z Z_+ + \right.\nonumber\\ &&\left.
\right.\nonumber\\ &&\left.(s_--s_+)_y Y_-\right]
\label{cross3}
\eea
where $s_\pm$ are the polarization vectors of $\tau^\pm$
in the proper reference frame.

In Ref.\cite{nos} it is shown that 1) the transverse
polarization term ($X_+$) is proportional to the real part of
the AWMM except for a
helicity-flip suppressed tree level background from the Standard Model,
2) the normal polarization term ($Y_-$) is proportional to the weak-electric
form factor $d^w_{\tau}$, and 3)  the longitudinal polarization term
($Z_+$) has the well
known Standard Model contribution plus a quantum correction given by the
real part of the AWMM.
The existence of an absorptive part in the weak-magnetic moment induces
 a new component $Y_+$ in the normal polarization.
These  components are:
\begin{eqnarray}
X_+&=&a\; \sin\theta_{\tau^-} \bigg\{-\left[2v^2+(v^2+a^2)\beta\cos
\theta_{\tau^-}\right]\frac{v}{\gamma s_wc_w} + \nonumber\\
& &\begin{array}{c} \\ \end{array}
 2\gamma\left[2v^2(2-\beta^2)+(v^2+a^2)\beta\cos
\theta_{\tau^-}\right] Re(a^w_{\tau})\bigg\}\label{x} \\
Y_+&=&\;-2 v \gamma \beta \sin\theta_{\tau^-}
\;[2 a^2+(v^2+a^2) \beta \cos \theta_{\tau^-}]\;
 Im(a_{\tau}^w) \label{y+}\\
Y_-&=&2a\gamma\beta\sin\theta_{\tau^-} \left[
2v^2+(v^2+a^2)\beta\cos\theta_{\tau^-}\right] (2 m_{\tau} d^w_{\tau}
/e) \label{y} \\
Z_+&=&-\frac{va}{s_wc_w}\left[(v^2+a^2)\beta(1+\cos^2\theta_{\tau^-})+
2(v^2+\beta^2a^2)\cos\theta_{\tau^-}\ \right]\nonumber\\
& &+2a\left[4v^2\cos\theta_{\tau^-}+
(v^2+a^2)\beta(1+\cos^2\theta_{\tau^-})\right] Re(a^w_{\tau})\label{z}
\end{eqnarray}
where $\alpha$ is the fine structure constant, $\Gamma_Z$ is
the $Z$-width, $\gamma=\frac{M_Z}{2 m_\tau}$,
$\beta=(1-\frac{1}{\gamma^2})^{\frac{1}{2}}$
are the dilation factor and $\tau$ velocity, respectively,
 and $d^w_{\tau}$
is the weak-electric form factor. The reference
frame is chosen  such that
the outgoing $\tau^-$ momenta is along
the $z$ axis and the incoming $e^-$ momenta is in the $x-z$ plane,
and $\theta_{\tau^-}$ is the angle determined by these two momenta.
Terms with $(s_--s_+)_{x,y,z}$ factors in Eq.(\ref{cross3})
 carry all the information
about the
CP violating pieces of the lagrangean.
The normal polarization is even under parity, then only
 $a\cdot v^2 \cdot d_{\tau}^w$
or $a^3 \cdot d_{\tau}^w$ terms are allowed in $Y_-$ (see Eq.(\ref{y})),
in contrast with the case in the spin-spin correlation observables, where the
leading term is  $a^2 \cdot v \cdot d_{\tau}^w$.
The new term with $(s_-+s_+)_y$ induced by the $Im(a_{\tau}^w)$
is CP-conserving and it is a time reversal-odd observable generated
by an absorptive part. The dependence with $a^2 \cdot v \cdot a_{\tau}^w$ or
$v^3 \cdot a_{\tau}^w$ is associated with the normal
polarization being even under parity.
Transverse polarization  of a single
$\tau$ (along the x-axis) is parity-odd
and time reversal-even, and it can only arise from the
interference of both helicity conserving and
helicity flipping amplitudes.
The first term of $X_+$ in Eq.(\ref{x}) comes from
helicity flipping suppressed
($\frac{1}{\gamma}\equiv \frac{2m_\tau}{M_Z}$)
 amplitudes in the Standard Model and the second one comes from
the $\gamma$-enhanced  chirality
flipping weak-magnetic tensorial $a_{\tau}^w$ vertex.

{}From Eqs.(\ref{cross1}) and (\ref{cross3})  the  $e^+ e^- \rightarrow
\tau^+\  \tau^-\rightarrow h^+_1 X\; h^-_2 \nu_\tau $ and
$h^+_1 \bar{\nu_\tau}\; h^-_2 X$ cross
 sections  \cite{alemany,tsai} are:

\begin{eqnarray}
&&\hspace{-1cm}\frac{d\sigma(e^+e^-\rightarrow \tau^+\tau^-
\rightarrow h^+_1 X h^-_2\nu_\tau)}
{d(\cos\theta_{\tau^-})\, d\phi_{h^-_2}}=
Br(\tau^-\rightarrow h^-_2\nu_\tau)Br(\tau^+
\rightarrow h^+_1
X) \times\nonumber\\
&&\hspace{2.cm}\left[
4 \frac{d\sigma^{0}}{d\Omega_{\tau^-}}+
\frac{\alpha^2\beta \pi}{128s_w^3 c_w^3\Gamma_Z^2}
\alpha_{h_2^-} (X_+\cos\phi_{h_2^-}+(Y_-+Y_+)\sin\phi_{h_2^-})\right]
\label{difcross1}\\
&&\hspace{-1cm}\frac{d\sigma(e^+e^-\rightarrow \tau^+\tau^-
\rightarrow h^+_1\bar{\nu_\tau}h^-_2 X)}
{d(\cos\theta_{\tau^-})\, d\phi_{h^+_1}}=
Br(\tau^-\rightarrow h^-_2
X)Br(\tau^+
\rightarrow h^+_1\bar{\nu_\tau}) \times\nonumber\\
&&\hspace{2.cm}\left[
4 \frac{d\sigma^{0}}{d\Omega_{\tau^-}}+
\frac{\alpha^2\beta\pi }{128s_w^3 c_w^3\Gamma_Z^2}
\alpha_{h_1^+}(-X_+\cos\phi_{h_1^+}+(Y_--Y_+)\sin\phi_{h_1^+})\right]
\label{difcross2}
\end{eqnarray}
where the angle $\phi_h$ is the
 azimuthal hadron angle in the frame we have already defined.
All other angles have been
integrated out.
The longitudinal polarization term ($Z_+$) disappears
when the polar angle $\theta_h$ of the hadron is
integrated out.
For $\pi$ and $\rho$ mesons the magnitude of the parameter
$\alpha_h$ is  $\alpha_\pi=0.97$ and  $\alpha_\rho=0.46$.
The spin correlation terms give no contribution to the
angular distributions Eqs.(\ref{difcross1}) and (\ref{difcross2}).

With the $\tau$ direction fully reconstructed \cite{kuhn}
in semileptonic
decays, as shown in Ref.\cite{nos}, we can get
 information about the AWMM, by
defining
  the following asymmetry of the
${\tau}$-decay products:
\begin{equation}
A_{cc}^\mp=\frac{\sigma^\mp_{cc}(+)-\sigma^\mp_{cc}(-)}
{\sigma^\mp_{cc}(+)+\sigma^\mp_{cc}(-)}
\end{equation}
with
\begin{equation}
\hspace{-.75cm}\sigma^\mp_{cc}(+)= \left[\int_{0}^{1}
d(\cos \theta_{\tau^-})
\int_{-\pi/2}^{\pi/2} d \phi_{h^\mp}
 + \int_{-1}^{0} d (\cos\theta_{\tau^-})
\int_{\pi/2}^{\frac{3}{2} \pi}d \phi_{h^\mp}\right]
\frac{d \sigma}{d (\cos\theta_{\tau^-}) \; d \phi_{h^\mp}}
\end{equation}
and
\begin{equation}
\hspace{-.75cm}\sigma^\mp_{cc}(-)= \left[\int_{0}^{1}
d (\cos\theta_{\tau^-})
\int_{\pi/2}^{\frac{3}{2}\pi} d \phi_{h^\mp} +
\int_{-1}^{0} d (\cos\theta_{\tau^-})
\int_{-\pi/2}^{\pi/2}d \phi_{h^\mp}\right]
\frac{d \sigma}{d (\cos\theta_{\tau^-})\; d \phi_{h^\mp}}
\end{equation}
This asymmetry selects the leading $\cos\theta_{\tau^-}\cos \phi_{h^\mp}$
term of the cross section.
After some algebra one finds
\begin{equation}
A_{cc}^\mp= \mp \alpha_h \frac{s_wc_w}{4 \beta}  \frac{v^2+a^2}{a^3}\left[
-\frac{v}{\gamma s_wc_w} +2\gamma \; Re(a^w_{\tau})\right]
\label{acc}\end{equation}
with opposite values for $\tau^-$ and $\tau^+$.

For numerical results we  consider $10^7 Z$ events and
one $\tau$ decaying into
$\pi \; \nu_\tau$ or  $\rho \;  \nu_\tau$
({\it i.e.} $h_1 , \;h_2= \pi$ or $\rho$ in
(\ref{difcross1}) and (\ref{difcross2}) respectively),
while  summing up over the $\pi \; \nu_\tau$, $\rho \;
  \nu_\tau$ and
$a_1 \; \nu_\tau$  semileptonic  decay channels
for the other $\tau$ (this amounts to
 about 52\% of the total decay rate).

Using  the asymmetry (\ref{acc})
  it is then possible to
measure the AWMM.
Collecting events from the decay of both {\it taus},
one gets a sensibility (within 1s.d.):
\begin{equation}
 \abs{Re(a_\tau^w)} \leq 4\cdot 10^{-4}
\label{awt}\end{equation}
for the combined data for $\pi$ and $\rho$ channels.

Let us discuss the possibility of measuring this magnitude using the
method outlined above.
There, the analysis was done assuming vanishing absorptive parts in
$a_{\tau}^w$. The real part  of
this magnitude appears in the transverse polarization.
The actual calculation shows that this is not a very good approximation,
and that both real and imaginary parts are of the same order  of magnitude.
We have shown  that the best
sensitivity one can expect is of the order of $10^{-4}$,
so the result (\ref{anom}) for
 the standard weak-magnetic moment will not be accessible
 in such an experiment.
 In order to disentangle the $a_{\tau}^w$ term one has to subtract
 the tree level helicity-flip term coming from the
Standard Model. This is the first
 term in the right hand side of Eq.(\ref{x}) and Eq.(\ref{acc}). Then, if an
 anomalously large signal related to the observable persists, it
should be attributed to  physics beyond the Standard Model.

While the real part of the AWMM contributes to the transverse
polarization, the absorptive part
of the AWMM  contributes to the CP-even terms of the
 normal polarization $Y_+$, and no mixing of this real and imaginary
 parts occurs in the polarization terms. The transverse polarization is then
  an observable related to the real part of the AWMM while the normal
  polarization (in the absence of a CP-violating interaction) is a magnitude
  related to the absorptive part. The CP-even part of the normal polarization
  $Y_+$ is given by Eq.(\ref{y+}).
  The analysis is then similar to the
 one  made in Ref.\cite{nos} for the weak-electric form factor.
 There,
 the leading $d_{\tau}^w$ term is extracted from an asymmetry that
picks up the $\sin\theta_{\tau} \cos\theta_{\tau} \sin \phi_{h}$  term.
However, in this case it is much better ( \ie, the sensitivity is enhanced
by a factor $3 \pi /4$)
to define an asymmetry that picks up the $\sin \phi_{h^\mp}$ term from $Y_+$:
\beq
{A_s}^\mp=
\frac{
\displaystyle{\int_{0}^{\pi} d \phi_h^{\mp}
\displaystyle{\frac{d\sigma}{d\phi_h^{\mp}}}}
-
\int_{\pi}^{2 \pi} d \phi_h^{\mp}
\displaystyle{\frac{d\sigma}{d\phi_h^{\mp}}}
}{\displaystyle{\int_{0}^{\pi} d \phi_h^{\mp}
\displaystyle{\frac{d\sigma}{d\phi_h^{\mp}}}}
+\int_{\pi}^{2 \pi} d \phi_h^{\mp}
\displaystyle{\frac{d\sigma}{d\phi_h^{\mp}}
}}\eeq
After some algebra one finds
\beq
{A_s}^\mp = \mp \alpha_h \frac{3 \pi\gamma}{4} c_w s_w
\frac{v}{a^2}\; Im(a_{\tau}^w)
\eeq

Under the same assumptions as for the real part, and collecting events for
both negative and positive {\it tau}-decays, we obtain that for
 the pion channel (\ie $h\equiv \pi$) it is possible to  put
the following bounds for the absorptive part of $a_{\tau}^w$:
 \beq
\abs{Im(a_{\tau}^w)} \leq 1.4 \times 10^{-3}\hspace{3cm} (\pi \;
{\rm channel})
\eeq
while for the $\rho$ channel
we have
\beq
\abs{Im(a_{\tau}^w)} \leq 2.0 \times 10^{-3} \hspace{3cm} (\rho \;
{\rm channel})
\eeq

Combining these results  one gets a sensibility:

\beq
\abs{Im(a_{\tau}^w)} \leq 1.1 \times 10^{-3}
\eeq

We have shown  that the best
sensitivity one can expect in the measurement of these observables
is of the order of $10^{-4}$,
so the result (\ref{anom}) for
the real and imaginary AWMM
 will not be accessible.
  Other proposals \cite{stiegler} have the same order of
magnitude for the sensitivity to  the real part of the AWMM
$a^w_\tau$. Thus a measurement
of the transverse and normal
polarization of single taus offers an opportunity to put bounds on
the weak moments induced by models beyond the standard theory.

\section*{Acknowledgements}

G.A.G.S. thanks the Spanish Ministerio de Educaci\'on y
Ciencia for a postdoctoral grant at the University of Valencia.
This work has been supported in part by CICYT, under
Grant AEN 93-0234, and by I.V.E.I..

\renewcommand{\thesection}{\Alph{section}}
\renewcommand{\theequation}{\thesection.\arabic{equation}}
\addtocounter{section}{-2}
\addtocounter{equation}{-22}
\section*{Appendix}

In this Appendix we explicitly show some of the formulas we have used
in the computation of the AWMM.

Current conservation for a vector current $V^\mu$ implies that
\bea
(p_-+p_+)_\mu \;\bar{u}(p_-) \,V^\mu(p_-,p_+) \,v(p_+) \,=\, 0
\label{cons}\eea
{}From the above equation we deduce that the form factor corresponding to
$(p_-+p_+)^\mu$ should vanish. Furthermore, the Gordon identity
\beq
\bar{u}(p_-)\, \gamma^\mu\, v(p_+)\, =\, \displaystyle{\frac{1}{2 m}}\,
 \bar{u}(p_-)\; [ i \sigma^{\mu\nu}
 (p_++p_-)_\nu+(p_--p_+)^\mu ]\; v(p_+)
\label{gordon}\eeq
leaves only two independent vector form factors:
the charge radius and the AWMM. When computing the loop integral
in any of the diagrams of figure 1 or figure 2, one naturally ends up with a
vector and axial tensor structure constructed with the only vectors
available, \ie the external
vectors $p_-$ and $p_+$, and the vectors and axial vectors
constructed entirely with the Gamma-matrices and the $\bar{u}$ and $v$ spinors.
  To extract the contribution to the
AWMM one has to select the $(p_--p_+)^\mu$ terms
from the tensor structure of the loop integrals, in
the basis of $\gamma^\mu$ and
$(p_--p_+)^\mu$ vectors.

 We define the scalar, vector and tensor one-loop 3-point functions as:
\bea
\{I_{00}\,;\,I_\mu\,;\,I_{\mu\nu}\}\; ({p_-}^2,(p_-+p_+)^2,{p_+}^2,{m_{_A}}^2,
{m_{_B}}^2,{m_{_C}}^2)\,=\, \displaystyle{
\frac{1}{i \pi^2}} \times \nonumber\\
\hspace*{-4cm}\int d^nk
 \displaystyle{
\frac{\{1\,;\,k_\mu\,;\,k_\mu k_\nu\}}{(k^2-{m_{_A}}^2)
((k-p_-)^2-{m_{_B}}^2)((k+p_+)^2-{m_{_C}}^2)}
}
\label{point}\eea
These functions have the following Lorentz structure:
\bea
I^\mu\,&=&\,({p_-}-{p_+})^\mu I_{10}+({p_-}+{p_+})^\mu I_{11} \nonumber\\
I^{\mu\nu}\,&=&\, ({p_+}^\mu {p_+}^\nu +{p_-}^\mu {p_-}^\nu) I_{21}+
({p_+}^\mu {p_-}^\nu+ {p_-}^\mu {p_+}^\nu) I_{22}+ \nonumber\\&&
({p_+}^\mu {p_+}^\nu -{p_-}^\mu {p_-}^\nu) I_{2-1}+
({p_+}^\mu {p_-}^\nu- {p_-}^\mu {p_+}^\nu) I_{2-2}+g^{\mu\nu} I_{20}
\label{3point}\eea
To extract the contribution to the AWMM one only needs to calculate the
$I_{00}$, $I_{10}$, $I_{21}$ and $I_{22}$ type integrals. Current conservation
and the structure of the contractions of the $I_{\mu\nu}$ integrals that
appears in the calculus (\ie , $I_{\eta\nu} g^{\eta\nu} \gamma^\mu$ and
$I_{\mu\nu} \gamma^\nu$) eliminates $I_{11}$, $I_{2-1}$, $I_{2-2}$ and
$I_{20}$.
These functions can be written in terms of analytic and dilogarithm
functions, as shown in \cite{pass}. Furthermore, one can also express all
of them  in terms  of the $I_{00}$ function and the 2-point functions.
To compute the actual value of the AWMM we have explicitly done
the calculus of these functions
for the external momenta on-shell and for the different
masses circulating in the loop,
and we have checked these results in the $m_\tau \rightarrow 0$ limit.
In the following we list the results for the $I_{ij}$ functions
(in the $m_\tau \rightarrow 0$ limit) needed
for the $a_{ABC}$ contributions given in Eq.(\ref{lead}), except for
$a_{\gamma\tau\tau}$  that it is proportional to the well known result for the
anomalous magnetic moment at $q^2=M_Z^2$.
As all external lines are on the mass shell, the arguments of the $I_{ij}$ are
given by the square of the masses of the particles in the corresponding
diagram.
For $a\ztautau$ they are:
\bea
{I_{00}}\,(m_\tau^2,M_Z^2,m_\tau^2,M_Z^2,0,0)\,=\,
\frac{1}{M_Z^2}\, \left( -\frac{\pi^2}{12} - i \pi ln2\right)\nonumber
\eea
\bea
 {I_{10}}\,(m_\tau^2,M_Z^2,m_\tau^2,M_Z^2,0,0)\,=\,
\frac{1}{M_Z^2}\, \left( - 1 + \frac{\pi^2}{12} +
i \pi ( ln2 - 1) \right)\nonumber
\eea\bea
{I_{21}}\,(m_\tau^2,M_Z^2,m_\tau^2,M_Z^2,0,0)\,=\,
\frac{1}{M_Z^2}\, \left( \frac{3}{4} - \frac{\pi^2}{12} + i \pi
\left( \frac{1}{2} - ln2 \right) \right)\nonumber
\eea
\bea
{I_{22}}\,(m_\tau^2,M_Z^2,m_\tau^2,M_Z^2,0,0)\,=\,
\frac{1}{M_Z^2}\, \left( - \frac{5}{2} + \frac{\pi^2}{4} - i \pi
( 2 -  3 ln3 ) \right)
\eea
For $a\nuww$, $a\nuws$ and $a\nusw$,
we define $r=\frac{M_W^2}{M_Z^2}$, and they are:
\bea
I_{00}\,(m_\tau^2,M_Z^2,m_\tau^2,0,M_W^2,M_W^2)
\,=&\,
\displaystyle{\frac{1}{M_Z^2}}\, \left(
\displaystyle{Li_2\left(\frac{r}{r-r_+}\right)-
Li_2\left(\frac{r-1}{r-r_+}\right)}+
 \right.\nonumber\\&\left. \hspace*{-5cm}
\displaystyle{Li_2\left(\frac{r}{r-r_-}\right)-
Li_2\left(\frac{r-1}{r-r_-}\right)+
Li_2\left(r\right)+ln(r)ln(1-r)-\frac{1}{2} ln^2(r)}\right)
\nonumber
\eea
\bea
{I_{10}}\,(m_\tau^2,M_Z^2,m_\tau^2,0,M_W^2,M_W^2)\,=\,
\displaystyle{\frac{1}{M_Z^2}\,
\left( 2 \sqrt{4 r-1} \,tan^{-1}
\left(\frac{1}{\sqrt{4 r-1}}\right)-1\right)} + r  I_{00}\nonumber
\eea
\bea
{I_{21}}\,(m_\tau^2,M_Z^2,m_\tau^2,0,M_W^2,M_W^2)\,=\,&
\displaystyle{\frac{1}{M_Z^2}}\, \left(- \displaystyle{\frac{3}{4}}
 - r +
 ( 1 + 2 r )\,\sqrt{4 r - 1} \,\times\right.\nonumber\\& \left. tan^{-1}
\left(\displaystyle{\frac{1}{\sqrt{4 r-1}}}\right) \right)+ r^2  I_{00}
\nonumber\eea
\bea
{I_{22}}\,(m_\tau^2,M_Z^2,m_\tau^2,0,M_W^2,M_W^2)\,=\,
- \frac{1}{2} \, \frac{1}{M_Z^2}  -  2 r  I_{10}
\eea
where $r_\pm=\frac{1\pm\sqrt{1-4 r}}{2}$ and $Li_2$ is the dilogarithm
function.
For $a\tfz$ and $a\tzf$,
when $M_\Phi = 2 M_Z$,  they are:
\bea
{I_{10}}\,(m_\tau^2,M_Z^2,m_\tau^2,0,M_Z^2,{M_\Phi}^2)\,=\,&
\displaystyle{\frac{1}{M_Z^2}\, \frac{1}{2}} \left(\displaystyle{
- 2 - \frac{2}{3} \pi^2 + 6 ln2 -4  i \pi ln\frac{3}{4}
- ln^22 } + \right.\nonumber\\&\left.
\displaystyle{4 Li_2\left( \frac{3}{4}\right)+ 8 Li_2\left( 2 \right)
 - 8 Li_2\left( \frac{3}{2}\right)}\right)
\eea
For $a\wnunu$, the functions are:
\bea
{I_{00}}\,(m_\tau^2,M_Z^2,m_\tau^2,M_W^2,0,0)\,=\,
\frac{1}{M_Z^2}\,\left( - ( i \pi + ln(r) ) \,ln\left(1+\frac{1}{r}\right)
 +Li_2\left(
-\frac{1}{r}\right)\right)\nonumber
\eea
\bea
{I_{10}}\,(m_\tau^2,M_Z^2,m_\tau^2,M_W^2,0,0)\,=\,&
\displaystyle{\frac{1}{M_Z^2}}\, \left( - 1 - ln(r) -
i \pi + ( i \pi + ln(r) ) r
ln\left(1+\displaystyle{\frac{1}{r}}\right) - \right.\nonumber\\
&\left. r Li_2\left(
-\displaystyle{\frac{1}{r}}\right)\right)\nonumber
\eea
\bea
{I_{21}}\,(m_\tau^2,M_Z^2,m_\tau^2,M_W^2,0,0)\,=\,&
\displaystyle{\frac{1}{M_Z^2}}\, \left( \displaystyle{
-\frac{1}{4} +  r + \left( r - \frac{1}{2} \right) ln(r) +
 i  \pi \left( r -\frac{1}{2} \right)} - \right.\nonumber\\&\left.
 ( i \pi + ln(r) ) \, r^2\, ln\left(1+\displaystyle{\frac{1}{r}}\right)
+ r^2 Li_2\left(
-\displaystyle{\frac{1}{r}}\right)\right)\nonumber
\eea
\bea
{I_{22}}\,(m_\tau^2,M_Z^2,m_\tau^2,M_W^2,0,0)\,=\,& \hspace*{-.2cm}
\displaystyle{\frac{1}{M_Z^2}}\, \left( \displaystyle{\frac{1}{2}}
 + 2 r + 2 r ln(r) + 2 i r^2 \pi
 + ( 1 + 2 r )
\,r \times \right.
\nonumber\\&\left. \,\left(\displaystyle{Li_2\left( -\frac{1}{r}\right)}
- ( i \pi + ln(r) ) \,ln\left(1+\displaystyle{\frac{1}{r}}\right)
 \right)\right)
\eea

\end{document}